\documentclass[preprint,prd,amsmath,amssymb,amsthm,nofootinbib]{revtex4}
\usepackage{xcolor}
\usepackage[mathscr]{euscript}
\usepackage{bm}
\usepackage{graphicx}
\usepackage[caption=false]{subfig}
\usepackage[colorlinks=true,linkcolor=red]{hyperref}
\usepackage{ulem}

\newcommand{\bea}{\begin{eqnarray}}
\newcommand{\eea}{\end{eqnarray}}
\newcommand{\beq}{\begin{equation}}
\newcommand{\eeq}{\end{equation}}

\begin{document}

\title{Geometric phase in AdS spacetime}
\author{Linghui Qiu$^{1}$, Jialin Zhang$^{1,2}$~\footnote{Corresponding author. jialinzhang@hunnu.edu.cn} and Hongwei Yu$^{1,2}$~\footnote{Corresponding author. hwyu@hunnu.edu.cn}}
\affiliation{
$^1${\small{Department of Physics, Key Laboratory of Low-Dimensional Quantum Structures and Quantum Control of Ministry of Education, Hunan Research Center of the Basic Discipline for Quantum Effects and Quantum Technologies, Hunan Normal University, 36 Lushan Rd., Changsha, Hunan 410081, China}}\\
$^2${\small{Institute of Interdisciplinary Studies, Hunan Normal University, 36 Lushan Rd., Changsha, Hunan 410081, China}}}

\begin{abstract}
We have investigated the geometric phase acquired by a uniformly accelerated Unruh-DeWitt detector  coupled to  vacuum fluctuations of a massless conformal  scalar field in anti-de Sitter (AdS) spacetime.  Using the open-quantum-system formalism, we calculate the phase under three  boundary conditions (Dirichlet, transparent, and Neumann) imposed on the field at the AdS boundary.  Our findings reveal a sharp distinction between subcritical and supercritical accelerations.  For subcritical accelerations, the detector evolves effectively as an isolated system, and the geometric phase is independent of both the AdS radius and the acceleration. For supercritical accelerations, however, topology-acceleration-induced phase corrections emerge and display pronounced boundary-condition dependence. When the AdS radius is smaller than the detector's proper wavelength, the magnitude of the correction at large accelerations follows the ordering Neumann$>$transparent$>$Dirichlet. Moreover,  over a finite interval of the detector's weight parameter, both Dirichlet and Neumann boundary conditions produce a richer peak structure in the phase correction than the transparent case,  with the detailed pattern  governed by the competition  between the acceleration and the detector's energy gap.
Finally, for transparent boundary conditions in the supercritical regime, the AdS phase correction closely resembles its de Sitter counterpart.

\end{abstract}

\maketitle

\section{Introduction}
The concept of a geometric phase dates back to Pancharatnam's study of interference between polarized light beams passing through crystals~\cite{Pancharatnam:1956}. In 1984, Berry discovered  that when a closed quantum system is transported adiabatically around a circuit in parameter space, its wave function acquires, in addition to the familiar dynamical phase, a purely geometric contribution~\cite{Berry:1984}. This idea was subsequently generalized to nonadiabatic cyclic processes~\cite{Aharonov:1987} and even to nonunitary, noncyclic processes~\cite{Samuel:1988}. In addition to theoretical advancements, experimental verification has been achieved using techniques  such as nuclear magnetic resonance interferometry~\cite{Du:2003} and single-photon interference~\cite{Ericsson:2005}.

In realistic settings, interaction with  environment is unavoidable, inevitably leading to decoherence and dissipation. Therefore, any real-world quantum system must be treated as an open quantum system, and consequently, the concept of the geometric phase must be extended accordingly~\cite{Uhlmann:1986,Sjoqvist:2000,Singh:2003,Ericsson:2003,Faria:2003,Tong:2004,ZSWang:2006}.   Environmental effects can modify the geometric phase in intriguing ways; notable examples include the influence of vacuum fluctuations~\cite{Lombardo:2006,Yu:2012,Cai:2018,Wang:2019,Villar:2020}, the Unruh effect~\cite{Martin-Martinez:2011,Hu:2012}, and the Hawking radiation~\cite{HuandYu:2012,Jing:2020}.

Building on the original proposal to detect the Unruh effect via the Berry (geometric) phase~\cite{Martin-Martinez:2011} and its extension to a
realistic Unruh-DeWitt (UDW) detector modeled by a  two-level atom~\cite{Hu:2012}, subsequent work has examined the geometric phase in a variety of settings.  For example, the geometric phase of a circularly accelerated atom in flat spacetime has been analyzed; for generic initial states the phase for circular motion differs from that of linear acceleration and can exceed it~\cite{Jin:2014}. Other studies have considered atoms undergoing linear or circular motion in free space or near reflecting boundaries and found that the geometric phase depends sensitively on the acceleration, the type of motion and the presence of a boundary~\cite{Zhai:2016,Zhao:2022}. The feasibility of detecting the acceleration effects inside a cavity using geometric phase was specifically discussed~\cite{Arya:2022}. In curved spacetimes, such as de Sitter (dS) space, freely falling and static atomic detectors acquire geometric-phase corrections that encode the Gibbons-Hawking temperature and the Unruh effect~\cite{Tian:2013}.

A natural question arising from these investigations is how a negative-curvature background such as anti-de Sitter (AdS) spacetime affects the geometric phase. AdS spacetime is maximally symmetric with constant negative curvature and plays a central role in string theory and cosmology. In particular, the AdS/CFT correspondence posits a duality between a gravitational theory in the bulk of AdS and a conformal field theory on its boundary~\cite{Maldacena:1999},  prompting intense interest in quantum fields on AdS backgrounds. AdS thus provides a valuable arena for studying quantum effects in curved spacetime and a test bed for relativistic phenomena.

 In this paper,  we examine a UDW detector undergoing uniform acceleration while interacting with a massless conformally coupled scalar field in the vacuum of AdS spacetime. We derive the geometric phase acquired by the detector and investigate how the Unruh effect and the spacetime curvature influence it. Our analysis treats three boundary conditions at spatial infinity, Dirichlet, transparent, and Neumann, and, through numerical evaluation, compares the results with the corresponding phases in dS spacetime.

The remainder of the paper is organized as follows. Section II reviews the basic equations governing the dynamical evolution of a UDW detector coupled to a  scalar field within  the open-quantum-system framework and derives a general formula for the  geometric phase. Section III presents the geometric phases of an accelerated UDW detector in AdS spacetime under three boundary conditions at spatial infinity (Dirichlet, transparent and Neumann) and compares them with those in dS spacetime via numerical analysis. Finally, we conclude with a summary in Sec. IV. Throughout this paper, we adopt the metric signature $(-, + ,+, +)$ and natural units $c=\hbar=k_B=1$.

\section{The basic formalism}
\label{sec2}
We consider a two-level quantum system (a UDW detector) coupled to a scalar field. The total Hamiltonian is the sum of three contributions:
\begin{equation}\label{pf1}
H=H_s+H_f+H_I\:.
\end{equation}
Here, $H_s$ is the Hamiltonian of the detector,
\begin{equation}
H_s=\frac{1}{2}\omega_{0}\big(|+\rangle\langle{+}|-|-\rangle\langle{-}|\big)=\frac{1}{2}\omega_{0}\sigma_3\;
\end{equation}
with $\omega_{0}$  the  energy gap between detector's excited state $|+\rangle$ and ground state $|-\rangle$ and $\sigma_i (i=1,2,3)$ are the Pauli matrices,
$H_f$ is the Hamiltonian of the free  field  $\phi(x)$, of which the explicit expression is not required here,  and the interaction Hamiltonian $H_I$ is
\begin{equation}
H_I=\mu(\sigma_{+}+\sigma_{-})\phi[x(\tau)]\;,
\end{equation}
where $\mu\ll1$ is small  coupling constant, and $\sigma_{+}=|+\rangle\langle{-}|$ and $\sigma_{-}=|-\rangle\langle{+}|$ denote the  raising and lowering operators of the UDW detector, respectively.
 The detector follows a classical trajectory  $x(\tau)$  parametrized by its proper time $\tau$.

We assume that at $\tau=0$ the combined system is in the product state  $\rho_{tot}(0)=\rho(0)\otimes|0\rangle\langle0|$, where $\rho(0)$ represents the initial density matrix of the detector and $|0\rangle$ is the  vacuum state of the field. The  evolution of the total system with respect to the proper time of the detector is governed by the von Neumann equation
\begin{equation}
\frac{\partial\rho_{tot}(\tau)}{\partial\tau}=-i[H,\rho_{tot}(\tau)]\;.\label{zfc1}
\end{equation}
Because the coupling is weak, we can trace over the field degrees of freedom and obtain a master equation for the detector's density matrix $\rho(\tau)$ of the Kossakowski-Lindblad form~\cite{Gorini:1976,Benatti:2003}
\begin{equation}
	\frac{\partial\rho(\tau)}{\partial\tau}=-i[H_{\text{eff}},\rho(\tau)]+\mathcal{L}[\rho(\tau)]\;,\label{zfc2}
\end{equation}
where $H_{\text{eff}}$ is the effective Hamiltonian of the detector, and $\mathcal{L}[\rho]$ is the  dissipator
\begin{equation}
	\mathcal{L}[\rho]=\frac{1}{2}\sum^3_{i,j=1}{a_{ij}(2\sigma_j \,\rho\,\sigma_i-\sigma_i\sigma_j\,\rho-\rho\,\sigma_i\sigma_j)}\;.\label{hsx}
\end{equation}	
In general, the effective Hamiltonian $H_{\text{eff}}$ and the Kossakowski matrix $a_{ij}$ in the dissipator are determined by the field Wightman function
\begin{equation}
	W\big(x(\tau),x'(\tau')\big):=\langle0|\phi(x(\tau))\phi(x'(\tau'))|0\rangle\,
\end{equation}
along the detector's worldline and its Fourier and Hilbert transforms. For a stationary trajectory of the detector, the corresponding Wightman function depends only on $\Delta\tau=\tau-\tau'$, and its  Fourier and Hilbert transforms  are  defined as follows:
\begin{equation}\label{Four-G}
	\mathcal{G}(\lambda)=\int_{-\infty}^{\infty}d\Delta\tau\, e^{i\lambda\Delta\tau}\,W(\Delta\tau)\,,~~~~\mathcal{K}(\lambda)=\frac{\mathcal{P}}{\pi i}\int_{-\infty}^{\infty}d\omega\,\frac{\mathcal{G}(\omega)}{\omega-\lambda}\,,
\end{equation}
where ${\mathcal{P}}$ represents the integral principal value.
After absorbing the Lamb shift contribution, the effective Hamiltonian $H_{\text{eff}}$ of the detector  then takes the form
\begin{equation}
	H_{\text{eff}}=\frac{1}{2}\Omega\sigma_3=\frac{1}{2}\Big\{\omega_{0}+\frac{i\mu^2}{2}
\Big[\mathcal{K}(-\omega_{0})-\mathcal{K}(\omega_{0})\Big]\Big\}\sigma_3\;,\label{Omega}
\end{equation}
while the coefficients of the Kossakowski matrix $a_{ij}$  can be expressed as
\begin{equation}
   a_{ij}=A\delta_{ij}-iB\epsilon_{ijk}\delta_{k3}-A\delta_{i3}\delta_{j3}\,,
\end{equation}
with
\begin{equation}
	A=\frac{\mu^2}{4}[\mathcal{G}(\omega_0)+\mathcal{G}(-\omega_0)]\,,~~B=\frac{\mu^2}{4}[\mathcal{G}(\omega_0)-\mathcal{G}(-\omega_0)]\;.\label{AB}
\end{equation}

It is convenient to write $\rho(\tau)$ in the Bloch representation:
\begin{equation}
	\rho(\tau)=\frac{1}{2}\big[1+\sum_{i=1}^{3}\rho_{i}(\tau)\sigma_{i}\big]\;
\end{equation}
with $(\rho_1,\rho_2,\rho_3)$ representing the Bloch vector.
Supposing  that the detector is initially prepared in a pure state, $|\psi(0)\rangle=\cos(\theta/2)|+\rangle+\sin(\theta/2)|-\rangle$
with $\theta$  being the weight parameter that characterizes the probability of the detector occupying each state, the master equation~(\ref{zfc2}) yields the exact solution
\begin{equation}
\left\{
\begin{aligned}
	\rho_1(\tau)&=e^{-2A\tau}\sin\theta\cos\Omega\tau\,,\\
	\rho_2(\tau)&=e^{-2A\tau}\sin\theta\sin\Omega\tau\,,\\
	\rho_3(\tau)&=e^{-4A\tau}\cos\theta+\frac{B}{A}(e^{-4A\tau}-1)\;.
\end{aligned}\right.
\end{equation}
From these components one reconstructs the time-dependent density matrix
\begin{equation}\label{m}
	\rho(\tau) = \Bigg(
	\begin{matrix}
		e^{-4A\tau}\cos^2\frac{\theta}{2}+\frac{B-A}{2A}(e^{-4A\tau}-1) & \frac{1}{2}e^{-2A\tau-i\Omega\tau}\sin\theta \\
		\frac{1}{2}e^{-2A\tau+i\Omega\tau}\sin\theta & 1-e^{-4A\tau}\cos^2\frac{\theta}{2}-\frac{B-A}{2A}(e^{-4A\tau}-1)
	\end{matrix}
	\Bigg)\;.
\end{equation}

For a mixed state undergoing nonunitary evolution, the geometric phase accumulated over a time interval $T$ can be defined as~\cite{Tong:2004}
\begin{equation}
\Phi=\arg\bigg(\sum^{N}_{k=1}\sqrt{\lambda_k(0)\lambda_k(T)}\langle\varphi_k(0)|\varphi_k(T)\rangle e^{-\int^{T}_{0}\langle\varphi_k(\tau)|\dot{\varphi}_k(\tau)\rangle d\tau}\bigg)\;,\label{geo}
\end{equation}
where $\lambda_k(\tau)$ and $|\varphi_k(\tau)\rangle$ are the eigenvalues and eigenvectors of the reduced density matrix $\rho(\tau)$.
In the present case, the eigenvalues are
\begin{equation}
	\lambda_{\pm}(\tau)=\frac{1}{2}(1\pm\eta)\label{e}\;,
\end{equation}
where $\eta=\sqrt{\rho_{3}^{2}(\tau)+e^{-4A\tau}\sin^2\theta}$.
Because $\lambda_{-}(0)=0$,  only the eigenvector $|\varphi_{+}(\tau)\rangle $ associated with $\lambda_{+}(\tau)$ contributes to  the geometric phase, and this vector can be expressed as
\begin{equation}\label{varphipos}
|\varphi_{+}(\tau)\rangle=\sin\frac{\theta_{\tau}}{2}|+\rangle+\cos\frac{\theta_{\tau}}{2}e^{i\Omega\tau}|-\rangle\,,
\end{equation}
where
\begin{equation}\label{thetatau}
	\sin\frac{\theta_{\tau}}{2}=\sqrt{\frac{\eta+\rho_3(\tau)}{2\eta}}\,,~~~~\cos\frac{\theta_{\tau}}{2}=\sqrt{\frac{\eta-\rho_3(\tau)}{2\eta}}\,.
\end{equation}
Substituting Eq.~(\ref{varphipos}) into Eq.~(\ref{geo}), one obtains  an expression for the geometric phase
\begin{align}\label{phig0}
	\Phi&=-\Omega\int_{0}^{T}\cos^{2}\frac{\theta_{\tau}}{2}d\tau=-\int_{0}^{T}\frac{\Omega}{2}\Big[1-\frac{R-R e^{4A\tau}+\cos\theta}{\sqrt{e^{4A\tau}\sin^{2}\theta+(R-R e^{4A\tau}+\cos\theta)^2}}\Big]{d}\tau\;
\end{align}
with $R:=B/A$.
 After performing the integral, the phase after one period  ($T=2\pi/\omega_0$),  can be written compactly in terms of an auxiliary function  $F(\varphi)$,
 \begin{equation}
	\Phi=\frac{\Omega}{\omega_{0}}[F(2\pi)-F(0)]\,,\label{phig2}
\end{equation}
where  $F(\varphi)$ is given by
\begin{equation}
\begin{aligned}
	F(\varphi)=&-\frac{1}{2}\varphi-\frac{\omega_{0}}{8A}\ln\Big[\frac{1-Q^2-R^2+2R^2e^{4A\varphi/\omega_0}}{2R}+S(\varphi)\Big]
\\&-\frac{\omega_0}{8A}\text{sgn}(Q)\ln\Big[1-Q^2-R^2+2Q^2e^{-4A\varphi/\omega_0}+2|Q|S(\varphi)e^{-4A\varphi/\omega_0}\Big]\,,
\end{aligned}
\end{equation}
with $S(\varphi)=\sqrt{R^2e^{8A\varphi/\omega_0}+(1-Q^2-R^2)e^{4A\varphi/\omega_0}+Q^2}$, $Q=R+\cos\theta$, and $\text{sgn}(Q)$ representing the  standard sign  function.

This formalism will be used in the next section to evaluate the geometric phase for an accelerated UDW detector interacting with a massless conformal scalar field in  AdS spacetime and to analyze how acceleration and spacetime curvature influence the phase.

\section{The geometric phase of an accelerated UDW detector in AdS spacetime}
\label{sec3}
Four-dimensional AdS spacetime can be represented as a hyperboloid embedded in five-dimensional flat spacetime with two timelike coordinates. In  Poincar\'e coordinates, the AdS metric can be expressed in the conformal flat form~\cite{Das:2001,Jennings:2010}
\begin{equation}
	ds^2=\frac{\ell^2}{z^2}( -dt^2+dx_1^2+dx_2^2+dz^2 )\;,
\end{equation}
where  $\ell$ represents the AdS radius and the negative cosmological constant is  $\Lambda=-3/\ell^2$. The  Poincar\'e boundary is at $z=0$, and there is a horizon at $z=\infty$ for  static observers.
In four-dimensional AdS spacetime, the  Wightman function of the massless conformal scalar field in its
vacuum state reads~\cite{Fronsdal:1974,Fronsdal:1975,Avis:1978,Deser:1997,Jennings:2010}
\begin{equation}
	W(x,x'):=\langle0|\phi(x)\phi(x')|0\rangle=\frac{1}{8\pi^2\ell^2}\bigg(\frac{1}{v-1}-\frac{\zeta}{v+1}\bigg)\,,\label{wightman}
\end{equation}
where
\begin{equation}
	v=\frac{z^2+z'^2+(\mathbf{x}-\mathbf{x}')^2-(t-t'-i\epsilon)^2}{2zz'}\,.\label{v}
\end{equation}
Since AdS space is not globally hyperbolic, one should impose  boundary conditions at spatial infinity ($z=0$) for   a well-defined field quantization scheme on four-dimensional AdS spacetime. Here, $\zeta$ specifies  boundary conditions: Dirichlet ($\zeta=1$), transparent ($\zeta=0$), and Neumann ($\zeta=-1$).

We now consider a UDW detector undergoing uniform acceleration through the AdS vacuum.  In AdS the magnitude of the proper acceleration $a$ relative to the inverse radius $1/\ell$ divides constant-acceleration orbits into three distinct classes~\cite{Bros:2002,Jennings:2010}:  elliptic (subcritical acceleration with $a<1/\ell$),  parabolic (critical acceleration with $a=1/\ell$) and hyperbolic trajectories (supercritical acceleration with $a>1/\ell$).  In the next section we will use the open-system formalism to compute the geometric phase acquired by a uniformly accelerated detector in AdS spacetime under Dirichlet, transparent and Neumann boundary conditions, and compare the results with those in dS spacetime.

\subsection{Subcritical accelerations}
When the magnitude of the proper acceleration satisfies $a\ell<1$, the detector follows a family of elliptic (subcritical) trajectories in AdS spacetime.   These worldlines can be parametrized in Poincar\'e  coordinates by~\cite{Das:2001,Jennings:2010}
\begin{equation}
	t(\tau)=\frac{A_0}{\ell}\sinh\gamma(\tau)\,,~~~~z(\tau)=\frac{A_0}{\ell}\cosh\gamma(\tau)+A_0a \,,
\end{equation}
with
\begin{equation}
	\cosh\gamma(\tau)=\frac{a\ell\cos(\sqrt{1/{\ell^2}-a^2}\tau)-1}{a\ell-\cos(\sqrt{1/{\ell^2}-a^2}\tau)}
\end{equation}
and  $A_0$ is a constant.
Along such a trajectory, the Wightman function of a massless conformally coupled scalar field depends only on the proper-time separation $\Delta\tau:=\tau-\tau'$.  For four-dimensional AdS, it takes the form
\begin{equation}\label{W0}
W_0(x,x')=\frac{1-a^2\ell ^2}{8\pi ^2\ell ^2}\Big[ \frac{1}{\cos ( \sqrt{1/\ell ^2-a^2}\Delta\tau -i\epsilon) -1}-\frac{\zeta}{\cos( \sqrt{1/\ell ^2-a^2}\Delta\tau -i\epsilon ) -2a^2\ell ^2+1} \Big]\;.
\end{equation}
 This Wightman function has no poles in the lower half-plane, so  ${\mathcal{G}}(-\omega_0)=0$. Therefore a uniformly accelerated detector moving along an elliptic trajectory  does not become excited and hence exhibits no Unruh-type thermal response~\cite{Jennings:2010}.    In this sense, the detector behaves effectively as if it were inertial, despite its nonzero acceleration.

 Because the detector sees no thermal bath, the deexcitation response ${\mathcal{G}}(\omega_0)$ can be calculated straightforwardly using Cauchy's residue theorem. The geometric phase accumulated over one period then becomes (see the Appendix)
 \begin{equation}
\Phi_{0}=-\pi(1-\cos\theta)\;.\label{phi00}
\end{equation}
This result is independent of the coupling constant, the AdS radius and the acceleration.
This is precisely the first term of the inertial Minkowski-space result~\cite{Hu:2012}
\begin{equation}\label{phim0}
\Phi_{M_0}\approx -\pi ( 1-\cos \theta) -\frac{\mu^2\pi}{4}( 2+\cos \theta)\sin ^2\theta\;.
\end{equation}
The difference
\begin{equation}\label{deltasub}
\delta=\Phi_{0}-\Phi_{M_0}\approx\frac{\mu^2\pi}{4}( 2+\cos \theta)\sin ^2\theta\;,
\end{equation}
is  precisely
the second term of $\Phi_{M_0}$  and arises from   the detector-field interaction in Minkowski spacetime. This  implies that the detector in subcritical accelerations evolves as an isolated quantum system, unaffected by the external environment.
For initial states $\theta=k\pi,~k\in\mathbf{Z}$ (the detector in a definite energy eigenstate), we find $|\Phi_{0}|=2\pi$ or $0$ and $\delta=0$. This reflects that for a detector prepared initially in its excited state $|+\rangle$  or ground state $|-\rangle$,  no geometric phase is accumulated in a complete cycle.

\subsection{Critical accelerations}
For a constant acceleration exactly equal to the inverse AdS radius  ($a\ell=1$), the detector follows the parabolic worldline. In Poincar\'e  coordinates, the trajectory can be parametrized as~\cite{Jennings:2010}
\begin{equation}\label{cr-ph}
t(\tau)=\frac{z_0\tau}{\ell},~~~~z(\tau)=z_0\;,
\end{equation}
where $z_0$ is a constant position. Substituting  this  trajectory  into Eq.~(\ref{wightman}) yields
\begin{equation}\label{G}
W_{c}(x,x')=-\frac{1}{4\pi^2}\bigg[\frac{1}{(\Delta\tau-i\epsilon)^2}-\frac{\zeta}{(\Delta\tau-i\epsilon)^2-4\ell^2}\bigg]\;.
\end{equation}
Its Fourier transform is therefore
\begin{equation}
    \begin{gathered}
        \mathcal{G}_{c}(\omega _0) =\frac{\omega _0}{2\pi}-\frac{\zeta\sin( 2\omega _0\ell)}{4\pi \ell}\,, \\
        \mathcal{G}_{c}(-\omega _0) =0\,.
    \end{gathered}\label{ab}
\end{equation}
Consequently, the  effective energy gap of the detector and the coefficients of the Kossakowski matrix  are given by

\begin{equation}\label{Omega-c}
	\Omega_{c}=\omega_0+\frac{\mu^2}{4\pi^2}{\mathcal{P}}\int_{0}^{\infty}d\omega\Big(\frac{1}{\omega+\omega_0}-\frac{1}{\omega-\omega_0}\Big)\Big(\omega-\frac{\zeta\sin(2\omega\ell)}
{2\ell}\Big)\;
\end{equation}
and
\begin{equation}
	A_{c}=B_{c}=\frac{\mu^2\omega_0}{8\pi}\bigg(1-\frac{\zeta\sin( 2\omega _0\ell )}{2\omega _0\ell}\bigg) \,.
\end{equation}

In the weak coupling  regime ($\mu^2\ll1$),   the Lamb shift term  in Eq.~(\ref{Omega-c}) can be neglected when
 computing  the  geometric phase caused by acceleration and AdS topology.
Using Eq.~(\ref{phig2}), the geometric phase in the critical case, to the order of $\mu^2$,  can then be approximated as
\begin{equation}
	\Phi_c \approx -\pi ( 1-\cos \theta) -\frac{\mu^2\pi}{4}\sin ^2\theta( 2+\cos \theta)\bigg( 1-\frac{\zeta\sin( 2\omega _0\ell)}{2\omega _0\ell}\bigg)\;.\label{go}
\end{equation}
In the flat-space limit $\ell\to\infty$,
this reduces to  the inertial Minkowski result,
\begin{equation}
	\lim_{\ell\rightarrow\infty}\Phi_c=	\Phi_{M_0}\approx -\pi ( 1-\cos \theta) -\frac{\mu^2\pi}{4}( 2+\cos \theta)\sin ^2\theta\,.
\end{equation}
Accordingly, the correction to the geometric phase arising purely from the influence of AdS topology and acceleration can be obtained as
\begin{equation}
	\delta=\Phi_c -\Phi_{M_0}\approx \frac{\zeta\mu^2\pi}{8\omega _0\ell}\sin( 2\omega _0\ell)(2+\cos \theta)\sin ^2\theta\;.\label{geoc}
\end{equation}
Evidently, $|\Phi_c|$ is equivalent to $0$ or $2\pi$ when $\theta=k\pi,~k\in\mathbf{Z}$, and the topology-acceleration-induced correction  $|\delta|$ vanishes identically.
Moreover, for a small AdS radius ($\omega_0\ell\ll1$), such topology-acceleration correction can be further approximated as
\begin{equation}\label{deltacr}
	\delta\approx \frac{\zeta\mu^2\pi}{4}(2+\cos \theta)\sin ^2\theta-\frac{\zeta\pi\mu^2\omega_0^2\ell^2}{6}(2+\cos \theta)\sin ^2\theta\;,
\end{equation}
where the leading term appears to match  Eq.~(\ref{deltasub})  in magnitude for both Dirichlet and Neumann boundary conditions.

\subsection{Supercritical accelerations}
For accelerations larger than the inverse AdS radius ($a\ell>1$), the constant-acceleration worldlines become hyperbolic.  In Poincar\'e  coordinates they can be written as
\begin{equation}
t(\tau)=\frac{a z_0}{\widetilde{\omega}_s}e^{\widetilde{\omega}_{s}\tau},~~~~z(\tau)=z_0e^{\widetilde{\omega}_s\tau}\;,
\end{equation}
where $z_0$ is  a constant and   $\widetilde{\omega}_{s}:=\sqrt{a^2-1/\ell^2}$. The Wightman function along these supercritical trajectories takes the form~\cite{Jennings:2010}
\begin{equation}
	W_{s}(x,x')=-\frac{\widetilde{\omega}_{s}^2}{16\pi ^2}\Big\{ \frac{1}{\sinh^2\big(\frac{\widetilde{\omega}_s
			\Delta \tau}{2}-i\epsilon\big)}-\frac{\zeta}{\sinh\big[\big(\frac{\widetilde{\omega}_s \Delta \tau}{2}-i\epsilon\big)+\widetilde{A}_\ell\big] \sinh\big[\big( \frac{\widetilde{\omega}_s \Delta \tau}{2}-i\epsilon\big)-\widetilde{A}_\ell\big]} \Big\}\,,	
\end{equation}	
where  $\widetilde{A}_\ell$  is defined by $\sinh \widetilde{A}_\ell:=\widetilde{\omega}_s\ell=\sqrt{a^2\ell^2-1}$ .

Using Eq.~(\ref{AB}), the coefficients of the Kossakowski matrix can be respectively written as
\begin{equation}
	\begin{aligned}
	&A_{s}=\frac{\mu^2\omega_0}{8\pi}\Big\{ 1-\frac{\zeta\sin \big[2\omega _0\ell\frac{{\rm{arcsinh}}(\sqrt{a^2\ell^2-1})}{\sqrt{a^2\ell^2-1}}
		 \big]}{2a\omega _0\ell^2} \Big\}  \coth\frac{\pi \omega _0\ell}{\sqrt{a^2\ell^2-1}}\,,\\
		&B_{s}=\frac{\mu^2\omega_0}{8\pi}\Big\{ 1-\frac{\zeta\sin \big[2\omega _0\ell\frac{{\rm{arcsinh}}(\sqrt{a^2\ell^2-1})}{\sqrt{a^2\ell^2-1}}
		 \big]}{2a\omega _0\ell^2} \Big\} \;.\label{AB1}
	\end{aligned}
\end{equation}
Substituting Eq.~(\ref{AB1}) into Eq.~(\ref{phig2}), we obtain  the geometric phase, to the order $\mu^2$
\begin{equation}
	\Phi_s\approx -\pi ( 1-\cos \theta ) -\frac{\mu^2\pi}
	{4}(N_s+1)\sin ^2\theta \Big(2+\cos \theta \coth \frac{\pi \omega _0\ell}{\sqrt{a^2\ell^2-1}}\Big)
\end{equation}
and the topology-acceleration-induced correction relative to the inertial Minkowski result becomes
\begin{equation}\label{geos}
	\delta=\Phi_s-\Phi_{M_0}\approx -\frac{\mu^2 \pi}{4}\sin ^2\theta\left[ \Big(2+\cos \theta \coth \frac{\pi \omega _0\ell}{\sqrt{a^2\ell^2-1}}\Big)N_s +\cos \theta \Big(\coth \frac{\pi \omega _0\ell}{\sqrt{a^2\ell^2-1}}-1\Big)\right]\;,
\end{equation}
where
\begin{equation}
	N_s= -\frac{\zeta}{2a\omega _0\ell^2}\sin \Big[ \frac{2\omega _0\ell}
	{\sqrt{a^2\ell^2-1}}{\rm{arcsinh}}(\sqrt{a^2\ell^2-1}) \Big]\,.
\end{equation}
Equation~(\ref{geos}) shows  that $\delta$ depends sensitively on the acceleration and the AdS radius. Specifically, in the limit of  $a\rightarrow1/\ell$,
the above expression continuously reduces to  Eq.~(\ref{geoc}),  recovering the critical-acceleration result.

For a small AdS radius compared with the detector's proper wavelength ($\ell\omega_0\ll1$), the correction to the geometric phase in Eq.~(\ref{geos}) can be further approximated as
\begin{equation}\label{small-adsr}
\delta\approx\left\{\begin{aligned}&
-\frac{\mu^2\pi\sin^2\theta}{4}\Big[\frac{\sqrt{a^2\ell^2-1}(1-\zeta)\cos\theta}{\ell\omega_0\pi}-2\zeta-\cos\theta\Big],&\ell\omega_0\ll\sqrt{a^2\ell^2-1}<1\;;\\\
&-\frac{\mu^2\pi\sin^2\theta\cos\theta}{4}\Big[\frac{a}{\pi\omega_0}-1-\frac{\zeta\ln(a\ell)}{a\pi\omega_0\ell^2}\Big],&\sqrt{a^2\ell^2-1}\gg1\gg\ell\omega_0\;.
        \end{aligned} \right.
\end{equation}
In contrast, for a large  AdS radius ($\ell\omega_0\gg1$), the approximation  is given by
\begin{equation}\label{large-adsr}
\delta\approx\left\{\begin{aligned}&
\frac{\mu^2\pi\sin^2\theta}{4}\Big[\frac{\zeta(2+\cos\theta)}{2\ell\omega_0}\sin\Big[\frac{\omega_0\ell(7-a^2\ell^2)}{3}\Big]
-{2\cos\theta}e^{-\frac{2\ell\omega_0\pi}{\sqrt{a^2\ell^2-1}}}
\Big],&\sqrt{a^2\ell^2-1}<1\ll\ell\omega_0\;;\\
&-\frac{\mu^2\pi\sin^2\theta\cos\theta}{2}\frac{1}{e^{2\pi\omega_0/a}-1},&\sqrt{a^2\ell^2-1}\gg\ell\omega_0\gg1\;.
        \end{aligned} \right.
\end{equation}

These limiting expressions make the parameter dependence transparent.
 For small AdS radius ($\ell\omega_0\ll1$), the magnitude  of the correction $\delta$ grows with acceleration when the acceleration far exceeds the inverse AdS radius [see last line of Eq.~(\ref{small-adsr})]. However, when the AdS radius exceeds the detector's proper wavelength ($\ell\omega_0\gg1$), the phase correction at moderate  acceleration [the first line of  Eq.~(\ref{large-adsr})] exhibits prominent oscillations as  $a\ell$ increases. In particular, in the extreme limit ($a\ell\gg1$), $\delta$  becomes independent of the boundary conditions [see the last line of  Eq.~(\ref{large-adsr})] and reduces to a purely thermal contribution.
 This follows from  the factor $({e^{2\pi\omega_0/a}-1})^{-1}$, which can be written as $({e^{\omega_0/T_{AU}}-1})^{-1}$  with  the AdS Unruh temperature $T_{AU}=\sqrt{a^2-\ell^{-2}}/(2\pi)$~\cite{Jennings:2010}. The corresponding phase shift is therefore of thermal origin~\cite{HuandYu:2012}.

To further elucidate the impact of the acceleration and AdS radius on the geometric phase, we now present numerical results for $\delta$.
\begin{figure}[!htbp]
	\centering
	\subfloat[$\theta=\pi/4,\ell\omega_0=0.1$]{\includegraphics[width=0.45\linewidth]{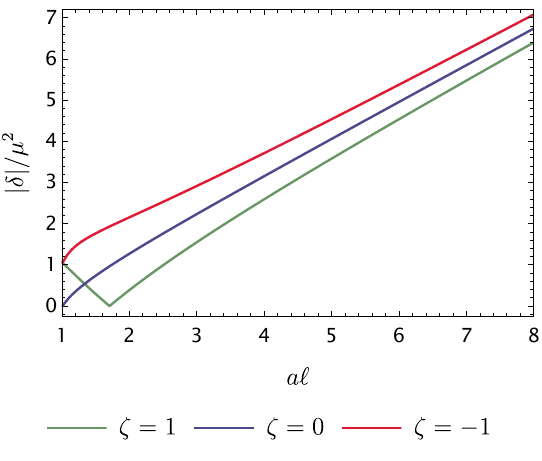}}~
	\subfloat[$\theta=\pi/4,\ell\omega_0=10$]{\includegraphics[width=0.45\linewidth]{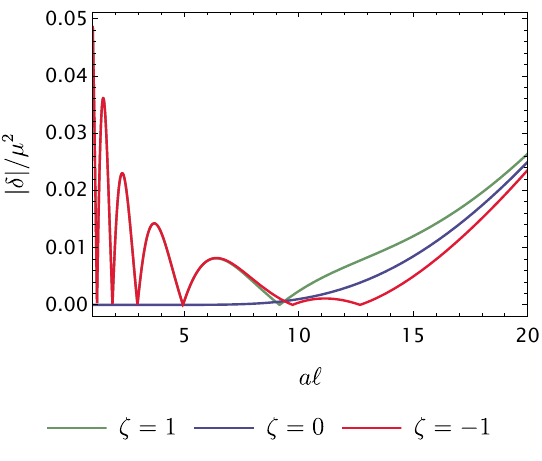}}
\caption{The magnitude of  the  topology-acceleration geometric-phase correction $|\delta|/\mu^2$ is plotted as a function of the parameter $a\ell$ ($a\ell\geq1$) in AdS spacetime with $\theta=\pi/4,\ell\omega_0=0.1$ in (a) and $\theta=\pi/4,\ell\omega_0=10$ in (b). Notice that for $\theta=\pi/4$ the subcritical acceleration case ($a\ell<1$) under all boundary conditions, according to Eq.~(\ref{deltasub}), then yields a constant phase correction : $|\delta|/\mu^2\approx1.06$.}\label{phasevsa}
\end{figure}
 Figure~(\ref{phasevsa}) shows $|\delta|/\mu^2$ as a function of $a\ell$  with $\theta=\pi/4,\ell\omega_0=0.1$ in Fig.~1(a) and $\theta=\pi/4,\ell\omega_0=10$ in Fig.~1(b).
  In the asymptotic regime  $a\ell\gg1$, $|\delta|$  increases monotonically with $a\ell$ for all the boundary conditions. Moreover, the discrepancies in $|\delta|$ among the three boundary conditions become significantly suppressed at large accelerations.
These trends are consistent with the analytical asymptotic forms in Eqs.~(\ref{small-adsr}) and~(\ref{large-adsr}).
For $\ell\omega_0\ll1$,  the ordering of the magnitude of $\delta$ at sufficiently large accelerations ($a\ell\gg1$) is Neumann $>$ transparent $>$Dirichlet. Conversely,  when  $\ell\omega_0\gg1$,  the Dirichlet and Neumann cases exhibit pronounced oscillations across moderately large $a\ell$.

\begin{figure}[!htbp]
	\centering
	\subfloat[$\ell\omega_0=0.5,a\ell=5$]{\label{phasevstheta11}\includegraphics[width=0.45\linewidth]{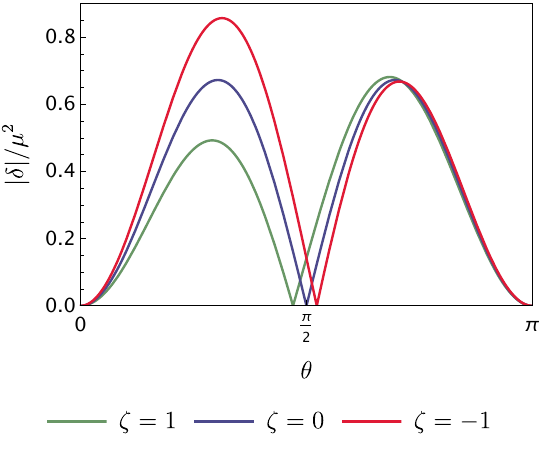}}~
	\subfloat[$\ell\omega_0=0.5,a\ell=10$]{\label{phasevstheta12}\includegraphics[width=0.45\linewidth]{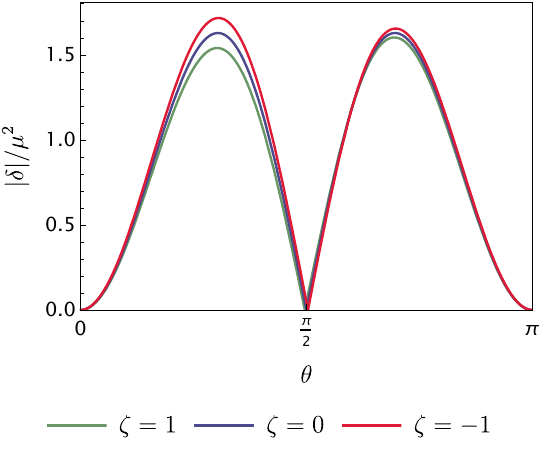}}\\
   \subfloat[$\ell\omega_0=5,a\ell=5$]{\label{phasevstheta21}\includegraphics[width=0.45\linewidth]{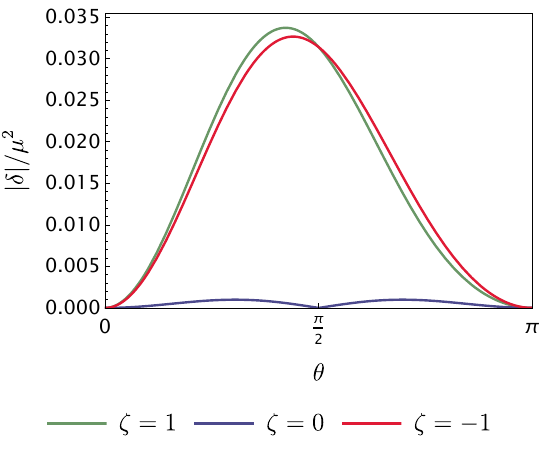}}~
	\subfloat[$\ell\omega_0=5,a\ell=10$]{\label{phasevstheta22}\includegraphics[width=0.45\linewidth]{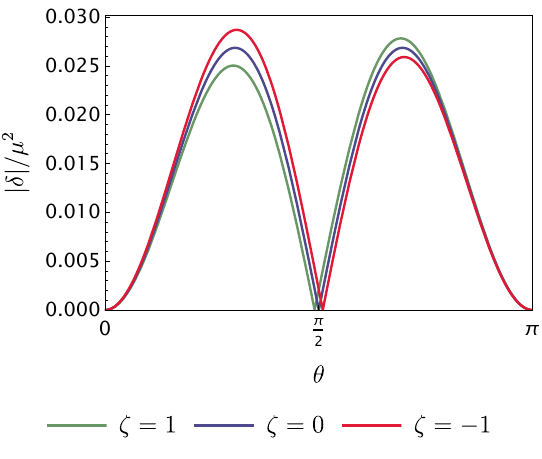}}
	\caption{The magnitude of the topology-acceleration geometric-phase correction versus the weight parameter for supercritical accelerations in AdS spacetime.  (a),(b) $\ell\omega_0=0.5$ with $a\ell=\{5,10\}$ in the left-to-right order. (c),(d) $\ell\omega_0=5$  with $a\ell=\{5,10\}$ in the left-to-right order.}\label{phasevstheta}
\end{figure}
Figure~(\ref{phasevstheta}) depicts the dependence of the geometric-phase correction $|\delta|$ on the weight parameter $\theta$ for supercritical accelerations under Dirichlet, transparent , and Neumann boundary conditions. As expected, $|\delta|$ vanishes at $\theta=0$  and $\theta=\pi$ for all boundaries. For intermediate $\theta$, the corrections are nonzero and differ markedly among these boundary conditions, although increasing $a\ell$  tends to reduce these discrepancies in $|\delta|$.

Notably, both Dirichlet and Neumann boundary conditions produce a more intricate peak structure in the phase correction than the transparent case. Specifically, they exhibit either one or two peaks within $\theta\in(0,\pi)$, depending sensitively  on the competition  between the acceleration $a$ and the energy gap $\omega_0$. As shown in  Figs.~(\ref{phasevstheta11}), (\ref{phasevstheta12}), and~(\ref{phasevstheta22}), when the detector's energy gap $\omega_0$ is much smaller than the acceleration $a$,  the  phase correction $|\delta|$  displays two peaks in $(0, \pi)$  for all three boundary conditions. This  follows from  the large-acceleration limit of Eq.~(\ref{geos}): $|\delta|\sim\sin^2\theta|\cos\theta|$ (independent of  all boundary choice), and the angular factor $\sin^2\theta|\cos\theta|$  has  two maxima within $\theta\in(0,\pi)$.

However, as the detector's energy gap becomes comparable to the acceleration, the $\theta$ dependence of $|\delta|$  develops distinct peak patterns: for both Dirichlet and  Neumann boundaries, the two-peak structure collapses to a single peak  near $\theta=\pi/2$;  whereas for transparent boundary conditions, two peaks persist [see Fig.~(\ref{phasevstheta21})].
This  is again consistent with  the large-gap estimate derived from  Eq.~(\ref{geos}).
For  $\omega_0>a$ and $\zeta\neq0$, the phase correction $|\delta|\sim\sin^2\theta(2+\cos\theta)|N_s|$, where  the factor $\sin^2\theta(2+\cos\theta)$ exhibits a single maximum near $\theta=\pi/2$ in  $(0,\pi)$. By contrast, for the transparent  case ($\zeta=0$), we still have $|\delta|\sim\sin^2\theta|\cos\theta|$, which produces two peaks in $(0,\pi)$. Therefore, as  long as the detector's energy gap is sufficiently large relative to the acceleration, a robust single-peak structure near $\theta=\pi/2$ is expected  for both Dirichlet and Neumann boundary conditions.

\subsection{Comparison with the dS spacetime}

Four-dimensional dS spacetime can, like AdS,  be realized as a hyperboloid embedded in five-dimensional Minkowski spacetime~\cite{Birrell:1982}. It is maximally symmetric but has constant positive curvature; the cosmological constant is $\Lambda=3/\ell^2$, where  $\ell$ denotes the dS radius.
A central role in inflationary cosmology is played by dS, whereas AdS is widely used in braneworld and gauge-gravity duality contexts.  Because the curvature signs are opposite, their global and thermodynamic properties differ markedly.

A comoving detector in dS spacetime perceives a thermal bath at the Gibbons-Hawking temperature $T_H=1/(2\pi\ell)$~\cite{Birrell:1982}. In general, a detector with constant proper acceleration $a$ in dS measures a temperature given by $T_{DU}=\sqrt{\ell^{-2}+a^2}/(2\pi)$~\cite{Narnhofer:1996}.

In the open-system approach, the geometric phase of a UDW detector undergoing nonunitary evolution depends sensitively on this thermal environment.  For a static detector in dS space (which experiences a constant proper acceleration determined by its radial position), the Wightman function leads to a Planckian spectrum with temperature $T_{DU}$. To leading order in the coupling strength $\mu$, the geometric phase over one cycle is~\cite{Tian:2013}
\begin{equation}
\Phi_{ds}\approx -\pi ( 1-\cos \theta ) -\frac{\mu^2 \pi}{4}\sin ^2\theta \big(2+\cos \theta+\frac{2\cos\theta}{e^{{2\pi\omega_0\ell}/{\sqrt{a^2\ell^2+1}}}-1}\big)\;.
\end{equation}
The last term arises from the thermal response: subtracting the flat-space inertial contribution $\Phi_{M_0}$ leaves a purely topology-and-acceleration-induced correction
\begin{equation}\label{deltads}
	\delta_{ds}=\Phi_{ds}-\Phi_{M_0}\approx -\frac{\mu^2 \pi}{2}\sin^2\theta\frac{\cos\theta}{e^{{2\pi\omega_0\ell}/{\sqrt{a^2\ell^2+1}}}-1}\;.
\end{equation}

Comparing with the AdS case, we find the following: First,  for  $a\ell<1$, an accelerated detector in AdS does not feel a thermal bath; the geometric phase reduces to the inertial result and is insensitive to curvature or acceleration.  In contrast, even an inertial detector in dS ($a=0$) sees a Gibbons-Hawking temperature.  Consequently, the  phase correction $\delta$ in dS is nonzero while its AdS counterpart vanishes. Second, for  $a\ell>1$, AdS detectors experience an Unruh-type temperature  $T_{AU}=\sqrt{a^2-\ell^{-2}}/(2\pi)$, and the geometric phase depends on boundary conditions (Dirichlet, transparent, or Neumann) as described in the preceding subsection.  In the high-acceleration limit, the phase correction becomes independent of the boundary condition and approaches a thermal form.

To demonstrate the distinct behavior of the geometric phases in AdS and dS spacetimes, we numerically compute the phase corrections attributable to the spacetime topology below.

\begin{figure}[!htbp]
	\centering
	\subfloat[$\theta=\pi/4,\ell\omega_0=0.2$]{\includegraphics[width=0.33\linewidth]{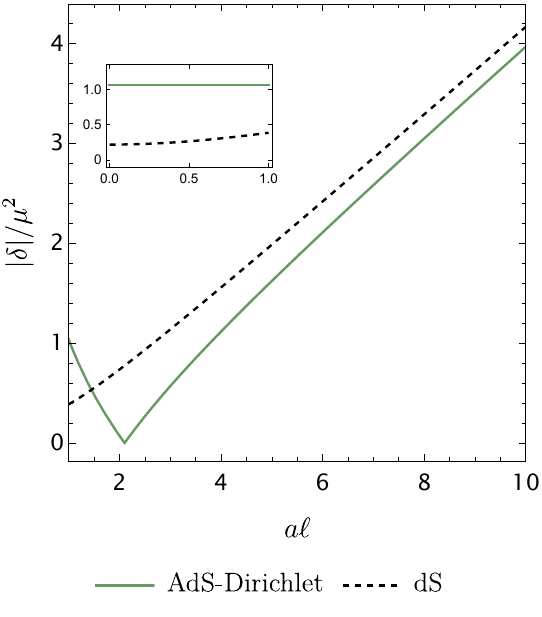}}~
	\subfloat[$\theta=\pi/4,\ell\omega_0=0.2$]{\includegraphics[width=0.33\linewidth]{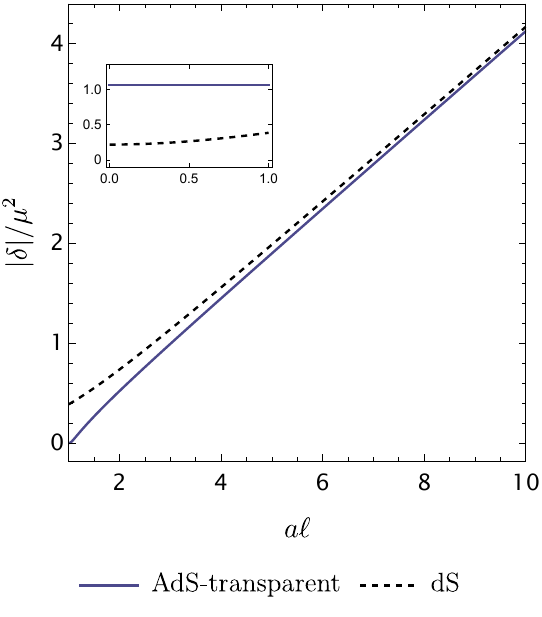}}~
\subfloat[$\theta=\pi/4,\ell\omega_0=0.2$]{\includegraphics[width=0.33\linewidth]{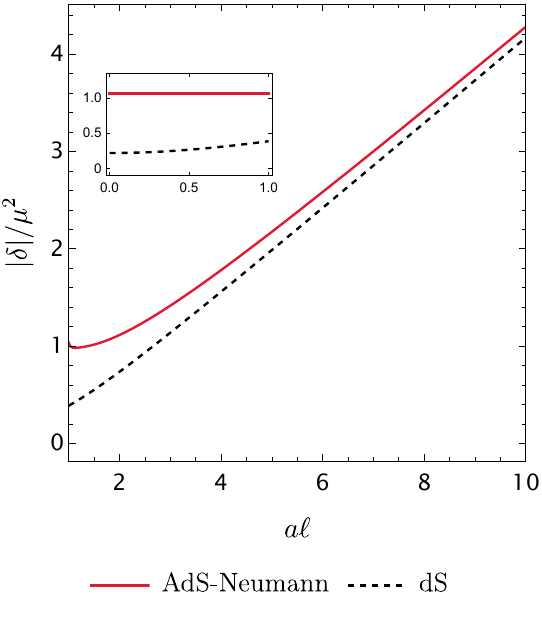}}
\vspace{0.2cm}
\subfloat[$\theta=\pi/4,\ell\omega_0=5$]{\includegraphics[width=0.33\linewidth]{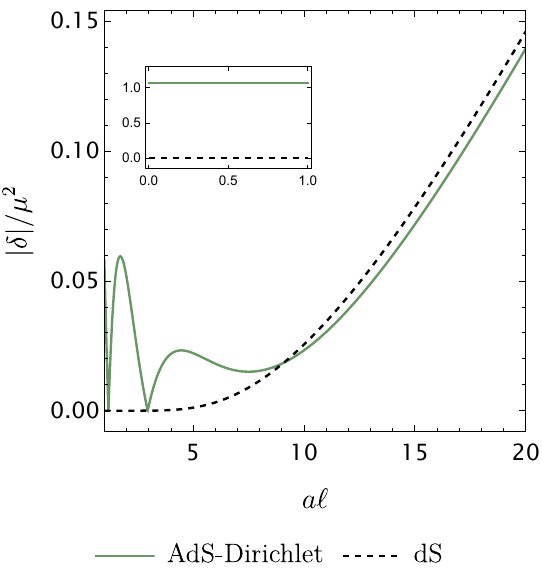}}~
	\subfloat[$\theta=\pi/4,\ell\omega_0=5$]{\includegraphics[width=0.33\linewidth]{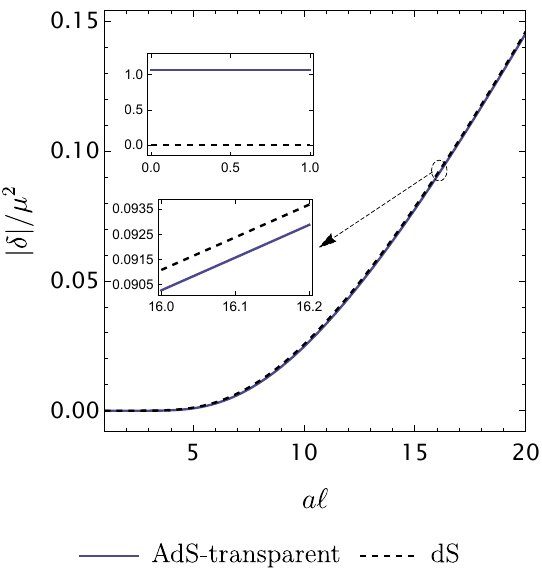}}~
\subfloat[$\theta=\pi/4,\ell\omega_0=5$]{\includegraphics[width=0.33\linewidth]{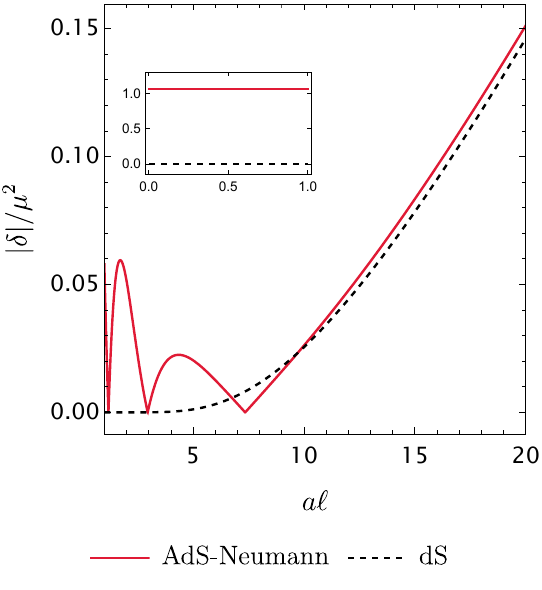}}
\caption{The topology-acceleration geometric-phase corrections  in AdS (under different boundary conditions)
 and dS spacetimes  are respectively plotted as a function of  $a\ell$ with $\theta=\pi/4,\ell\omega_0=0.2$  in (a)-(c) and with $\theta=\pi/4,\ell\omega_0=5$ in(d)-(f).
 }\label{com1}
\end{figure}
 Figure~(\ref{com1}) shows that  for small accelerations ($a\ell<1$) the phase corrections in dS and AdS differ significantly because dS has an intrinsic thermal bath while AdS does not~\cite{Jennings:2010,Birrell:1982,Narnhofer:1996,Casadio:2011}.  As the acceleration increases ($a\ell\gg1$), the corrections in both spacetimes converge; the dS result aligns closely with the AdS result under transparent boundary conditions, with the AdS corrections under Dirichlet and Neumann conditions bracketing the dS curve. The effective temperature $T_{DU}$  is generally larger than $T_{AU}$, and the resulting phase correction remains slightly larger in dS than in AdS for Dirichlet boundaries but smaller than in AdS with Neumann boundaries.

\begin{figure}[!htbp]
	\centering
	\subfloat[$\ell\omega_0=1,a\ell=1.5$]{\includegraphics[width=0.3\linewidth]{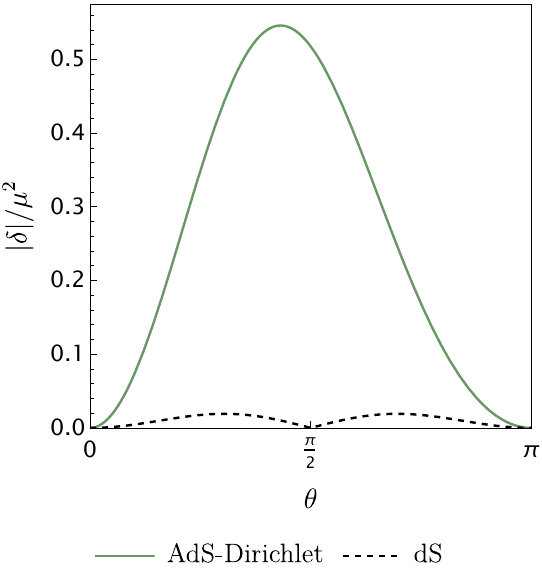}}~
	\subfloat[$\ell\omega_0=1,a\ell=1.5$]{\includegraphics[width=0.3\linewidth]{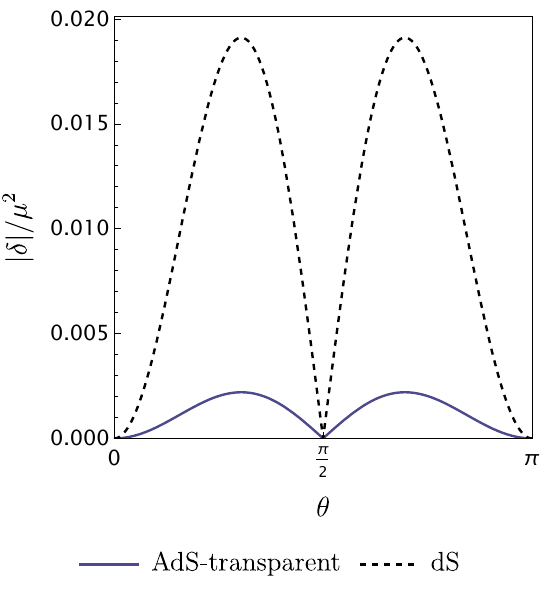}}~
\subfloat[$\ell\omega_0=1,a\ell=1.5$]{\includegraphics[width=0.3\linewidth]{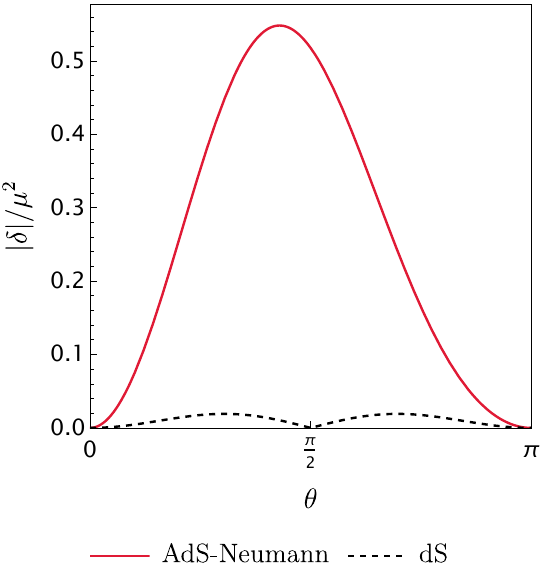}}
\vspace{0.2cm}
	\subfloat[$\ell\omega_0=1,a\ell=10$]{\includegraphics[width=0.3\linewidth]{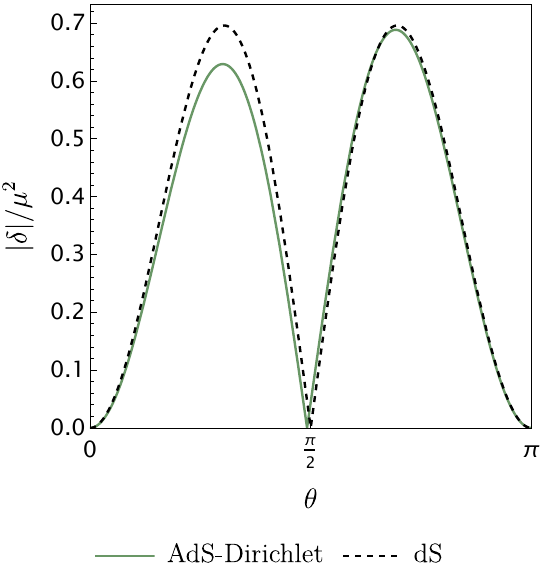}}~
	\subfloat[$\ell\omega_0=1,a\ell=10$]{\includegraphics[width=0.3\linewidth]{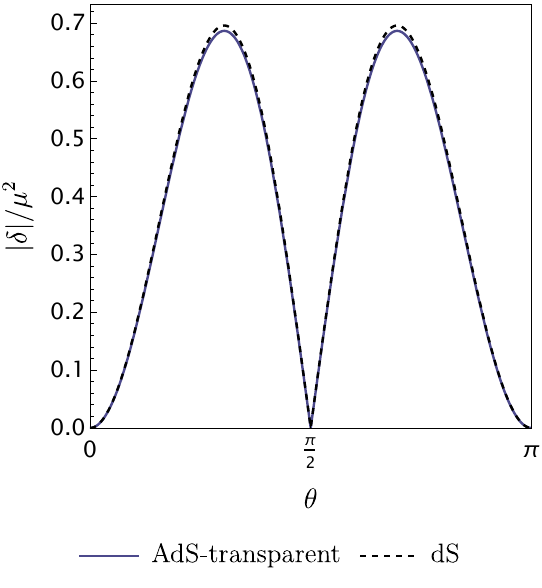}}~
\subfloat[$\ell\omega_0=1,a\ell=10$]{\includegraphics[width=0.3\linewidth]{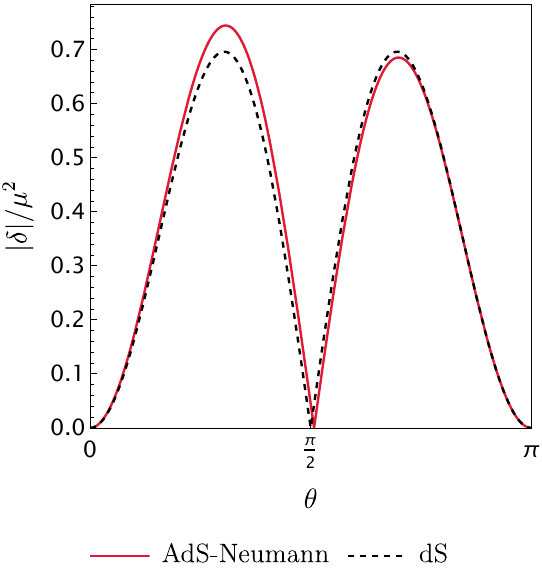}}
\caption{The topology-acceleration geometric-phase corrections in AdS (under different boundary conditions) and dS spacetimes  versus the parameter with $a\ell=1.5$ in (a)-(c) and $a\ell=10$ in (d)-(f). Here, we assume $\ell\omega_0=1$ for all plots.}\label{com2}
\end{figure}
In Fig.~(\ref{com2}),  we plot the magnitude of the topology-acceleration correction $|\delta|$ as a function of the initial-state parameter $\theta$  for both dS space and AdS space with different boundary conditions. The results show that the behavior of $|\delta|$ with respect to weight parameter $\theta$ in dS spacetime is analogous to that in AdS spacetime under transparent boundary conditions, exhibiting two  peaks located
over the interval $(0,\pi)$. In contrast, the double-peak structures of $|\delta|$ over $(0,\pi)$  for Dirichlet and Neumann boundary conditions require an additional condition, $a\gg\omega_0$. Overall, the discrepancy in the detector's geometric phases between dS and AdS spacetimes can be significantly reduced by increasing the acceleration.

 In summary, the geometric phase of an accelerated UDW detector acts as a sensitive probe of the thermal properties of dS and AdS spacetimes.   In dS space the phase correction always reflects the intrinsic Gibbons-Hawking radiation; in AdS space it arises only above a critical acceleration and depends on the boundary condition.
At very high accelerations, the differences between these two curved spacetimes diminish,  but at low accelerations the quasithermal nature of AdS and the truly thermal nature of dS lead to qualitatively distinct geometric phases. Consequently, these characteristic topology-acceleration phase corrections provide a valuable probe for the Unruh-type effect in curved spacetimes.

\section{Conclusion}
\label{sec5}

We have investigated how the geometric phase of a uniformly accelerated UDW detector is influenced by the curvature and topology of AdS spacetime.  Using the open-quantum-system framework, we derived the phase acquired by the detector when it interacts with a massless conformal scalar field, taking into account three types of boundary conditions (Dirichlet, transparent, and Neumann) at the AdS boundary.

Our analysis reveals a clear dichotomy between subcritical and supercritical accelerations: for subcritical accelerations, the detector does not perceive a thermal bath; its dynamical evolution and geometric phase are indistinguishable from those of an isolated system.  Consequently, the geometric phase is independent of both the AdS radius $\ell$ and the acceleration $a$.  In contrast, once the proper acceleration exceeds the curvature scale, the detector experiences an Unruh-like thermal effect in AdS.  The resulting geometric phase depends on the boundary condition and exhibits rich behavior.  At very high accelerations and for a small AdS radius ($\ell\omega_0 \ll 1$), the magnitude of the phase correction follows the hierarchy: Neumann $>$ transparent $>$ Dirichlet.   When the AdS radius is large compared with the detector's wavelength ($\ell\omega_0 \gg 1$), the corrections for Dirichlet and Neumann boundaries show oscillatory dependence on the product $a\ell$. Moreover, over a finite range of the detector's weight parameter, Dirichlet and Neumann boundary conditions display a more intricate peak structure than the transparent case, with the pattern governed by the competition between the acceleration and the energy gap.

Comparing AdS with dS spacetime further underscores how curvature sign alters the physics.  In dS space a comoving detector always sees a thermal bath at the Gibbons-Hawking temperature, and a constantly accelerated detector measures a combined temperature $T_{DU}=\sqrt{\ell^{-2}+a^2}/(2\pi)$.  In AdS, by contrast, only accelerations above the critical value $a=1/\ell$ produce a temperature $T_{\mathrm{AU}} = \sqrt{a^2 - \ell^{-2}}/(2\pi)$, and the geometric phase becomes independent of $a$ and $\ell$ below this threshold.  For large accelerations, the AdS phase corrections converge toward the dS result under transparent boundary conditions, whereas the Dirichlet and Neumann cases bracket the dS values.  Moreover, as the acceleration becomes very large, the AdS-dS differences in the phase correction become progressively suppressed.  These contrasts highlight the role of spacetime topology in quantum phases and suggest that geometric phases could serve as probes of Unruh-type effects in curved backgrounds.

Finally, the methodology and insights developed here invite extensions to other settings.  It would be natural to investigate the geometric phase of accelerated detectors near black holes or in other curved spacetimes.  Such studies could deepen our understanding of the interplay between quantum phases, acceleration and curvature.

\begin{acknowledgments}
This work was supported in part by the NSFC under Grants No.~12175062 and No.~12075084 and the innovative research group of Hunan Province under Grant No. 2024JJ1006.
\end{acknowledgments}

\appendix
\section{The geometric phase for subcritical accelerations}\label{Derivation-sub}
Substituting Eq.~(\ref{W0}) into the Fourier transform expression~(\ref{Four-G}), we obtain
\begin{align}\label{G01}
\mathcal{G}(\omega_0)&=\frac{1-a^2\ell ^2}{8\pi ^2\ell ^2}\int_{-\infty}^{\infty} \frac{e^{i\omega_0\Delta\tau}}{\cos ( \sqrt{1/\ell ^2-a^2}\Delta\tau -i\epsilon) -1}d\Delta\tau\nonumber\\
&\quad - \frac{1-a^2\ell ^2}{8\pi ^2\ell ^2}\int_{-\infty}^{\infty} \frac{\zeta e^{i\omega_0\Delta\tau}}{\cos( \sqrt{1/\ell ^2-a^2}\Delta\tau -i\epsilon ) -2a^2\ell ^2+1}d\Delta\tau\,.
\end{align}

Since $\omega_0>0$, only the poles in the upper half-plane contribute to the above integrals.
For the first integral in Eq.~(\ref{G01}),  the poles in the upper half-plane are located at
\begin{equation}
\Delta\tau=\frac{2n\pi}{\sqrt{1/\ell^2-a^2}}+i \epsilon\,,\quad n \in \mathbf{Z}\,
\end{equation}
with $a<1/\ell$.
Here $\epsilon$  is a small positive regulator,  whose precise magnitude is irrelevant since any positive prefactor can be absorbed. Using the residue theorem, the contour integral becomes
\begin{align}\label{adint-1}
&\frac{1-a^2\ell ^2}{8\pi ^2\ell ^2}\int_{-\infty}^{\infty} \frac{e^{i\omega_0\Delta\tau}}{\cos ( \sqrt{1/\ell ^2-a^2}\Delta\tau -i\epsilon) -1}d\Delta\tau
\nonumber\\&=\lim_{\epsilon\to0^+}\Big[\sum_{n=1}^{\infty}\frac{\omega_0}{2\pi}e^{-\frac{2i n\pi\omega_0}{\sqrt{1/\ell ^2-a^2}}-n\epsilon}+\sum_{n=1}^{\infty}\frac{\omega_0}{2\pi}e^{\frac{2i n\pi\omega_0}{\sqrt{1/\ell ^2-a^2}}-n\epsilon}+\frac{\omega_0}{2\pi}\Big]
\nonumber\\&=0\;.
\end{align}

For the second integral in  Eq.~(\ref{G01}), the poles in the upper half-plane are
\begin{equation}
\Delta\tau = \frac{\pm \arccos(2a^2 \ell^2 - 1) + 2n\pi }{\sqrt{1/\ell^2 - a^2}}+ i\epsilon,\quad n \in \mathbf{Z}\,.
\end{equation}
A similar residue calculation gives
\begin{align}\label{adint-2}
&-\frac{1-a^2\ell ^2}{8\pi ^2\ell ^2}\int_{-\infty}^{\infty} \frac{\zeta e^{i\omega_0\Delta\tau}}{\cos( \sqrt{1/\ell ^2-a^2}\Delta\tau -i\epsilon ) -2a^2\ell ^2+1}d\Delta\tau
\nonumber\\=&\frac{\zeta}{8\pi a\ell^2} \lim_{\epsilon \to 0^{+}}\Big\{\sum_{n=1}^{\infty}ie^{\frac{i\omega_0[-2n\pi+\arccos(2a^2\ell^2-1)]}{\sqrt{1/\ell^2-a^2}}-n\epsilon}-\sum_{n=1}^{\infty}ie^{\frac{i\omega_0[-2n\pi-\arccos(2a^2\ell^2-1)]}{\sqrt{1/\ell^2-a^2}}-n\epsilon}\nonumber
\\&+\sum_{n=1}^{\infty}ie^{\frac{i\omega_0[2n\pi+\arccos(2a^2\ell^2-1)]}{\sqrt{1/\ell^2-a^2}}-n\epsilon}-\sum_{n=1}^{\infty}ie^{\frac{i\omega_0[2n\pi-\arccos(2a^2\ell^2-1)]}{\sqrt{1/\ell^2-a^2}}-n\epsilon}-2\sin\Big[\frac{\omega_0\arccos(2a^2\ell^2-1)}{\sqrt{1/\ell^2-a^2}}\Big] \Big\}\nonumber
\\=&0\;.
\end{align}

Combining Eq.~(\ref{adint-1}) and  Eq.~(\ref{adint-2})  yields ${\mathcal{G}}(\omega_0)=0$ for subcritical accelerations. Since $\mathcal{G}(-\omega _0) =0 $ as well,  there is neither excitation nor deexcitation response for subcritical accelerations. It follows that  $A=B=0$, $\Omega=\omega_0$, and consequently the dissipator ${\mathcal{L}}(\rho)=0$. The master equation therefore reduces to purely unitary evolution, with the exact solution
\begin{equation}
\left\{
\begin{aligned}
	\rho_1(\tau)&=\sin\theta\cos\omega_0\tau\,,\\
	\rho_2(\tau)&=\sin\theta\sin\omega_0\tau\,,\\
	\rho_3(\tau)&=\cos\theta\;.
\end{aligned}\right.
\end{equation}
The corresponding density matrix is
\begin{equation}\label{A-den}
	\rho(\tau) = \Bigg(
	\begin{matrix}
		\cos^2\frac{\theta}{2} & \frac{1}{2}e^{-i\omega_0\tau}\sin\theta \\
		\frac{1}{2}e^{i\omega_0\tau}\sin\theta & \sin^2\frac{\theta}{2}
	\end{matrix}
	\Bigg)\;.
\end{equation}

The eigenvector  $|\varphi_{+}(\tau)\rangle $  associated with nonzero eigenvalue $\lambda_{+}(\tau)$ of the density matrix~(\ref{A-den}) can then be written as
\begin{equation}
|\varphi_{+}(\tau)\rangle:=\sin\frac{\theta_{\tau}}{2}|+\rangle+\cos\frac{\theta_{\tau}}{2}e^{i\omega_0\tau}|-\rangle=\cos\frac{\theta}{2}|+\rangle+\sin\frac{\theta}{2}e^{i\omega_0\tau}|-\rangle\;.
\end{equation}
Using the geometric-phase formula in Eq.~(\ref{phig0}), we have
\begin{equation}
\Phi_{0}=-\omega_0\int_{0}^{T}\cos^{2}\frac{\theta_{\tau}}{2}d\tau=-\omega_0\int_{0}^{\frac{2\pi}{\omega_0}}\sin^{2}\frac{\theta}{2}d\tau=-\frac{\omega_0}{2}\int_{0}^{\frac{2\pi}{\omega_0}}(1-\cos\theta)d\tau=-\pi(1-\cos\theta)\;.
\end{equation}
This is in exact agreement with the general expression for the geometric phase, Eq.~(\ref{phig0}), evaluated under the conditions $A=0$ and  $\Omega=\omega_0$.



\begin{thebibliography}{}
\bibitem{Pancharatnam:1956}
S. Pancharatnam, 
\href{https://doi.org/10.1007/BF03046050}
{Proc. Indian Acad. Sci. A {\bf44}, 247 (1956).}

\bibitem{Berry:1984}
M. V. Berry, 
\href{https://doi.org/10.1098/rspa.1984.0023}
{Proc. R. Soc. Lond. A {\bf392}, 45 (1984).}

\bibitem{Aharonov:1987}
Y. Aharonov and J. Anandan, 
\href{https://doi.org/10.1103/PhysRevLett.58.1593}
{Phys. Rev. Lett. {\bf58}, 1593 (1987).}

\bibitem{Samuel:1988}
J. Samuel and R. Bhandari, 
\href{https://doi.org/10.1103/PhysRevLett.60.2339}
{Phys. Rev. Lett. {\bf60}, 2339 (1988).}

\bibitem{Du:2003}
J. Du, P. Zou, M. Shi, L. C. Kwek, J.-W. Pan, C. H. Oh, A. Ekert, D. K. L. Oi, and M. Ericsson, 
\href{https://doi.org/10.1103/PhysRevLett.91.100403}
{Phys. Rev. Lett. {\bf91}, 100403 (2003).}

\bibitem{Ericsson:2005}
M. Ericsson, D. Achilles, J. T. Barreiro, D. Branning, N. A. Peters, and P. G. Kwiat, 
\href{https://doi.org/10.1103/PhysRevLett.94.050401}
{Phys. Rev. Lett. {\bf94}, 050401 (2005).}

\bibitem{Uhlmann:1986}
A. Uhlmann, 
\href{https://doi.org/10.1016/0034-4877(86)90055-8}
{Rep. Math. Phys. {\bf24}, 229 (1986).}

\bibitem{Sjoqvist:2000}
E. Sj\"oqvist, A. K. Pati, A. Ekert, J. S. Anandan, M. Ericsson, D. K. L. Oi, and V. Vedral, 
\href{https://doi.org/10.1103/PhysRevLett.85.2845}
{Phys. Rev. Lett. {\bf85}, 2845 (2000).}

\bibitem{Singh:2003}
 K. Singh, D. M. Tong, K. Basu, J. L. Chen, and J. F. Du, 
\href{https://doi.org/10.1103/PhysRevA.67.032106}
{Phys. Rev. A {\bf67}, 032106 (2003).}

\bibitem{Ericsson:2003}
M.~Ericsson, E.~Sj{\"o}qvist, J.~Br{\"a}nnlund, D.~K.~L.~Oi, and A.~K.~Pati, 
\href{https://doi.org/10.1103/PhysRevA.67.020101}
{Phys. Rev. A {\bf67}, 020101 (2003).}

\bibitem{Faria:2003}
J. G. Peixoto de Faria, A. F. R. de Toledo Piza, and M. C. Nemes,
\href{https://doi.org/10.1209/epl/i2003-00440-4}
{Europhys. Lett. {\bf62}, 782 (2003).}

\bibitem{Tong:2004}
D. M. Tong, E. Sj\"oqvist, L. C. Kwek, and C. H. Oh, 
\href{https://doi.org/10.1103/PhysRevLett.93.080405}
{Phys. Rev. Lett. {\bf93}, 080405 (2004).}

\bibitem{ZSWang:2006}
Z. S. Wang, L. C. Kwek, C. H. Lai, and C. H. Oh, 
\href{https://doi.org/10.1209/epl/i2006-10057-1}
{Europhys. Lett. {\bf74}, 958 (2006).}

\bibitem{Lombardo:2006}
F. C. Lombardo and P. I. Villar, 
\href{https://doi.org/10.1103/PhysRevA.74.042311}
{Phys. Rev. A   {\bf74}, 042311 (2006).}

\bibitem{Yu:2012}
H. Yu and J. Hu, 
\href{https://doi.org/10.1103/PhysRevA.86.064103}
{Phys. Rev. A {\bf86}, 064103 (2012).}

\bibitem{Cai:2018}
H. Cai and Z. Ren, 
\href{https://doi.org/10.1088/1361-6382/aaba64}
{Class. Quantum Grav. {\bf35}, 105014 (2018).}

\bibitem{Wang:2019}
Z. Wang and C. Xu, 
\href{https://doi.org/10.1209/0295-5075/126/50005}
{Europhys. Lett. {\bf126}, 50005 (2019).}

\bibitem{Villar:2020}
L. Viotti, F. C. Lombardo, and P. I. Villar, 
\href{https://doi.org/10.1103/PhysRevA.101.032337}
{Phys. Rev. A {\bf101}, 032337 (2020).}

\bibitem{Martin-Martinez:2011}
E. Mart\'{i}n-Mart\'{i}nez, I. Fuentes, and R. B. Mann, 
\href{https://doi.org/10.1103/PhysRevLett.107.131301}
{Phys. Rev. Lett. {\bf107}, 131301 (2011).}

\bibitem{Hu:2012}
J. Hu and H. Yu, 
\href{https://doi.org/10.1103/PhysRevA.85.032105}
{Phys. Rev. A {\bf85}, 032105 (2012).}

\bibitem{HuandYu:2012}
J. Hu and H. Yu, 
\href{https://doi.org/10.1007/JHEP09(2012)062}
{J. High Energy Phys. {09} (2012) 062.}

\bibitem{Jing:2020}
J. Jing, Z. Cao, X. Liu, and Z. Tian, 
\href{https://doi.org/10.1088/1361-6382/ab6b6d}
{Class. Quantum Grav. {\bf37}, 085001 (2020).}

\bibitem{Jin:2014}
Y. Jin, J. Hu, and H. Yu, 
\href{https://doi.org/10.1103/PhysRevA.89.064101}
{Phys. Rev. A {\bf89}, 064101 (2014).}


\bibitem{Zhai:2016}
H. Zhai, J. Zhang, and H. Yu, 
\href{http://dx.doi.org/10.1016/j.aop.2016.05.017}
{Ann. Phys. {\bf371}, 338 (2016).}

\bibitem{Zhao:2022}
Z. Zhao and B. Yang, 
\href{https://doi.org/10.1103/PhysRevD.106.036013}
{Phys. Rev. D {\bf106}, 036013 (2022).}

\bibitem{Arya:2022}
N. Arya, V.  Mittal, K. Lochan, and S. K. Goyal, 
\href{https://doi.org/10.1103/PhysRevD.106.045011}
{Phys. Rev. D {\bf106}, 045011 (2022).}


\bibitem{Tian:2013}
Z. Tian and J. Jing, 
\href{https://doi.org/10.1007/JHEP04(2013)109}
{J. High Energy Phys. {04} (2013) 109.}



\bibitem{Maldacena:1999}
J. Maldacena, 
\href{https://doi.org/10.1023/A:1026654312961}
{Int. J. Theor. Phys. {\bf38}, 1113 (1999).}

\bibitem{Gorini:1976}
V. Gorini, A. Kossakowski, and E. C. G. Sudarshan, 
\href{https://doi.org/10.1063/1.522979}
{J. Math. Phys. {\bf17}, 821 (1976)}; G. Lindblad, 
\href{https://doi.org/10.1007/BF01608499}
{Commun. Math. Phys. {\bf48}, 119 (1976).}

\bibitem{Benatti:2003}
F. Benatti, R. Floreanini, and M. Piani, 
\href{https://doi.org/10.1103/PhysRevLett.91.070402}
{Phys. Rev. Lett. {\bf91}, 070402 (2003).}

\bibitem{Das:2001}
S. R. Das and A. Zelnikov, 
 \href{https://doi.org/10.1103/PhysRevD.64.104001}
{Phys. Rev. D {\bf64}, 104001 (2001).}

\bibitem{Jennings:2010}
D. Jennings, 
\href{https://doi.org/10.1088/0264-9381/27/20/205005}
{Class. Quantum Grav. {\bf27}, 205005 (2010).}

\bibitem{Fronsdal:1974}
C. Fronsdal, 
\href{https://doi.org/10.1103/PhysRevD.10.589}
{Phys. Rev. D {\bf10}, 589 (1974).}

\bibitem{Fronsdal:1975}
C. Fronsdal and R. B. Haugen, 
\href{https://doi.org/10.1103/PhysRevD.12.3810}
{Phys. Rev. D {\bf12}, 3810 (1975).}

\bibitem{Avis:1978}
S. J. Avis, C. J. Isham, and D. Storey, 
\href{https://doi.org/10.1103/PhysRevD.18.3565}
{Phys. Rev. D {\bf18}, 3565 (1978).}

\bibitem{Deser:1997}
S. Deser and O. Levin, 
\href{https://doi.org/10.1088/0264-9381/14/9/003}
{Class. Quantum Grav. {\bf14}, L163 (1997).}



\bibitem{Bros:2002}
J. Bros, H. Epstein, and U. Moschella , 
\href{https://doi.org/10.1007/s00220-002-0726-z}
{Commun. Math. Phys. {\bf231}, 481 (2002).}


\bibitem{Birrell:1982}
N. D. Birrell and  P. Davies, Quantum Fields in Curved Space (Cambridge University Press, Cambridge , 1982).


\bibitem{Narnhofer:1996}
H. Narnhofer, I. Peter, and W. Thirring, 
\href{https://doi.org/10.1142/S0217979296000611}
{Int. J. Mod. Phys. B {\bf10}, 1507 (1996).}

\bibitem{Casadio:2011}
R. Casadio, S. Chiodini, A. Orlandi, G. Acquaviva, R. Di Criscienzo, and L. Vanzo, 
\href{https://doi.org/10.1142/S0217732311036516}
{Mod. Phys. Lett. A {\bf26}, 2149 (2011).}




\end{thebibliography}
\end{document}